\documentclass[twocolumn, superscriptaddress, nofootinbib, amsmath, amssymb,
    aps, prd, floatfix]{revtex4-2}
\usepackage{graphicx}
\usepackage{epsfig}  
\usepackage[dvipsnames]{xcolor}
\usepackage{setspace}
\usepackage{color}
\usepackage{epsf}    
\usepackage{amsfonts}
\usepackage{filecontents}
\usepackage{float}
\usepackage{subfig}
\usepackage{mwe}
\usepackage{dcolumn}
\usepackage{bm}
\usepackage{dcolumn}
\usepackage{textcomp}
\usepackage{multirow}
\usepackage{slashed}
\usepackage{float}
\restylefloat{figure}
\usepackage{adjustbox}
\usepackage[]{hyperref}
  \hypersetup{
  bookmarks=true,         
  unicode=true,         
  pdftoolbar=true,     
  pdfmenubar=true,     
  pdffitwindow=true,    
  pdfstartview={FitH},    
  pdfsubject={Sterilenus@DUNE},  
  pdfnewwindow=true,     
  pdfcreator={RevTeX},
  colorlinks=true,     
  linkcolor=red,   
  citecolor=blue,    
  filecolor=black,   
  urlcolor=blue,           
  }

\usepackage{csquotes}
 
\usepackage{soul}
\definecolor{mygreen}{rgb}{.1,.75,.1}

\newcommand{\rom}[1]{\uppercase\expandafter{\romannumeral #1\relax}}

\usepackage{hypcap}

\usepackage{CJK}

\begin{document}
\title{Reconstructing masses for semi-invisibly decaying particles pair-produced at lepton colliders}

\author{Jin Min Yang}
\email[]{jmyang@itp.ac.cn}
\affiliation{CAS Key Laboratory of Theoretical Physics, 
Institute of Theoretical Physics, Chinese Academy of Sciences, Beijing 100190, P. R. China}
\affiliation{School of Physical Sciences, University of Chinese Academy of Sciences, Beijing 100049, P. R. China}

\author{Yang Zhang}
\email[]{zhangyangphy@zzu.edu.cn}
\affiliation{School of Physics, Zhengzhou University, Zhengzhou 450000, P. R. China}
\affiliation{CAS Key Laboratory of Theoretical Physics, 
Institute of Theoretical Physics, Chinese Academy of Sciences, Beijing 100190, P. R. China}

\author{Pengxuan Zhu}
\email{zhupx99@icloud.com}
\affiliation{CAS Key Laboratory of Theoretical Physics, Institute of Theoretical Physics, Chinese Academy of Sciences, Beijing 100190, P. R. China}

\author{Rui Zhu}
\email{zhurui@itp.ac.cn}
\affiliation{CAS Key Laboratory of Theoretical Physics, 
Institute of Theoretical Physics, Chinese Academy of Sciences, Beijing 100190, P. R. China}
\affiliation{School of Physical Sciences, University of Chinese Academy of Sciences, Beijing 100049, P. R. China}
\date{\today}

\begin{abstract} 

We present a set of Lorentz invariant kinematic variables for reconstructing mass of semi-invisible decaying particles pair-produced at lepton colliders, $m_{\rm RC}^{\rm min}$, $m_{\rm RC}^{\rm max}$ and $m_{\rm LSP}^{\rm max}$, with analytical formulas. They give the minimal and maximum bounds of the decaying particle mass and upper bound of the invisible particle mass. In the search of new physics, these variables can greatly enhance the statistical significance of signal. For the process of smuon pair production at $\sqrt{s}=240~{\rm GeV}$ lepton collider of 5 ab$^{-1}$, the cross section detection limit is pushed by one order, and the expected exclusion and discovery limits are set above $\sqrt{s}\big/2$ and go into the off-shell region. Moreover, these variables can also be used to improve the precision of $W$-boson mass measurement in full leptonic decayed channel. At future lepton collider, the precision can reach to $2\sim 3$ MeV level.
  
\end{abstract}
\maketitle

{\bf Introduction}~~~~ 
To account for dark matter (DM), almost all the extensions of the Standard Model (SM) contain additional particles charged under some new symmetry and/or possess an exact parity. In these theories, the lightest particle is stable and serves as the DM candidate. These theories can be tested at colliders via pair production of new particles which decay into invisible particles (missing energy) in detectors. Reconstructing the masses of such produced new particles is challenging since the energy carried away by the invisible particles can not be directly measured. 
 
At hadron colliders, inspired by this issue, new variables, such as $m_{\rm T2}$~\cite{Lester:1999tx, Barr:2003rg, Cheng:2008hk}, and new techniques, such as the recursive jigsaw reconstruction~\cite{Jackson:2016mfb, Jackson:2017gcy}, were developed and widely used. These techniques should be promoted to lepton colliders (LCs), as various machines like ILC~\cite{ Behnke:2013xla, Bambade:2019fyw, ILCInternationalDevelopmentTeam:2022izu}, FCC-ee~\cite{FCC:2018byv, Agapov:2022bhm}, CEPC~\cite{CEPCStudyGroup:2018ghi, Ruan:2018yrh, CEPCPhysicsStudyGroup:2022uwl} and Muon Collider~\cite{Black:2022cth, Delahaye:2019omf, Long:2020wfp} are being proposed for precise measurements. These state-of-the-art experiments enable us to scrutinize the nature of the Higgs boson~\cite{Li:2010wu, Haddad:2014fma, CEPCStudyGroup:2018ghi} and shed new light on new physics~\cite{Athron:2022uzz,Han:2020uak}.

\begin{figure}[t]
	\centering
    	\includegraphics[width=0.9\linewidth]{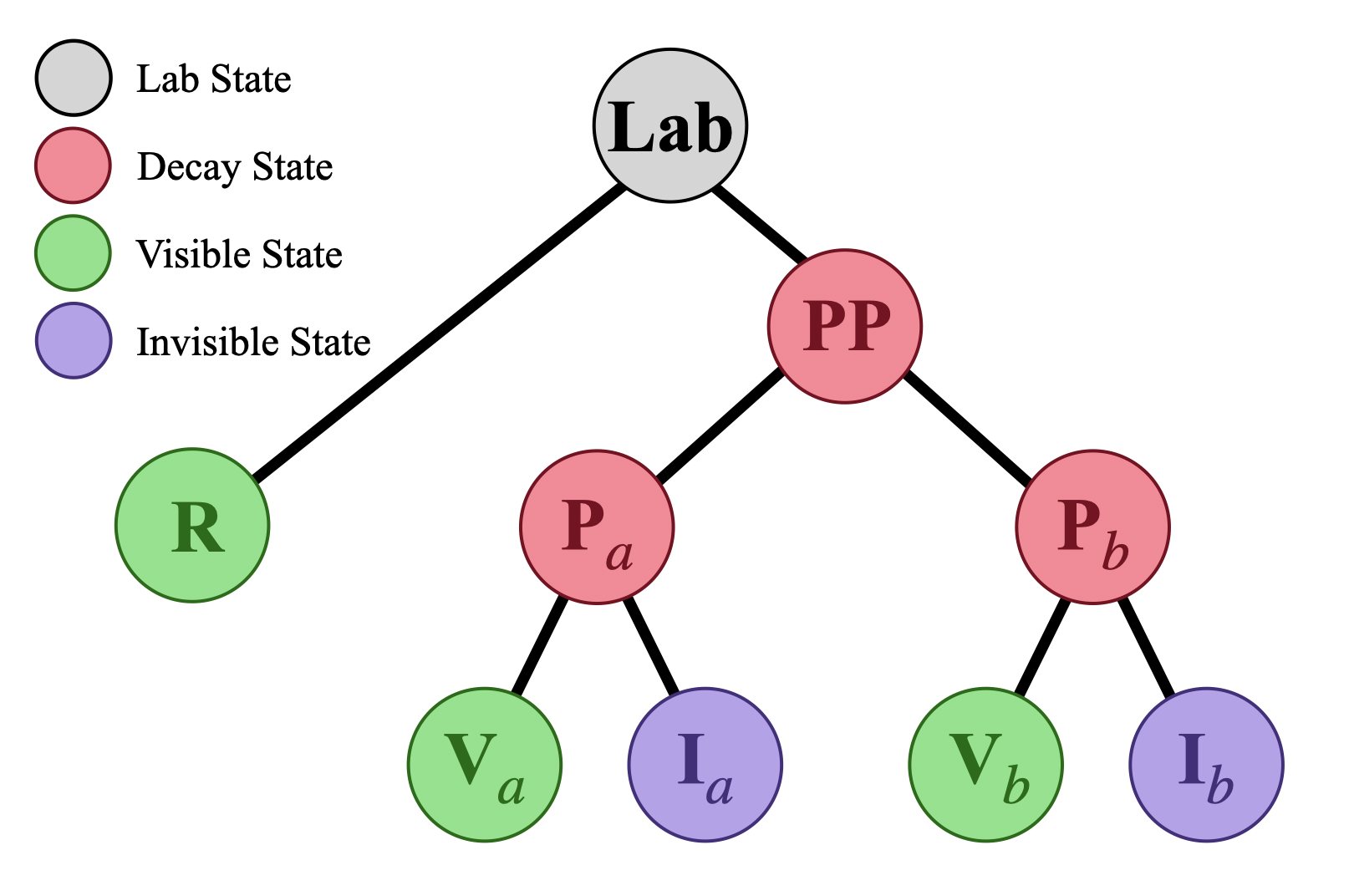}
\caption{\label{fig1} The decay tree of our considered process.}
\end{figure} 
\vspace{.5cm}

{\bf New kinematic variables}~~~~
In this work we introduce a set of kinematic variables, namely $m_{\rm LSP}^{\rm max}$, $m_{\rm RC}^{\rm max}$ and $m_{\rm RC}^{\rm min}$,  for a generic process shown in FIG.~\ref{fig1}, where a leptonic collision produces a pair of massive particles (${\bf P}{\bf P}$) with ${\bf P}$ decaying into an (a group of) observed particle(s) ${\bf V}$ and an invisible particle ${\bf I}$. Here we restrict ourselves to a symmetric decay tree, so the labels $a$ and $b$ in the diagram just indicate a same particle ${\bf P}$ in different branches.   

\par Throughout this paper, $p_{\bf a}^{\rm F}$ represents the four-momentum vector of object $\bf a$, evaluated in reference frame $\rm F$. The energy and momentum are denoted by $p_{\bf a}^{\rm F} \equiv \left(E_{\bf a}^{\rm F}, \vec{p}_{\bf a}^{\rm F} \right)$. If $\rm F$ is not declared, then the energy and/or momentum quantity is defined in the c.m. frame of $\bf PP$ system.

For an given event of the process depicted in FIG.~\ref{fig1}, we can read out the following four observables at a LC: $p_{{\bf V}_a}^{\rm Lab}$, $p_{{\bf V}_b}^{\rm Lab}$, $p_{\rm miss}^{\rm Lab}$ and $p_{\bf R}^{\rm Lab}$. The $\rm Lab$ represents the laboratory frame. $p_{\rm miss}^{\rm Lab}$ is the sum of the four-momentum vectors of all detector invisibly particles, and is given as follows
	\begin{equation}\label{eq:pmiss}\begin{split}
			p_{\rm miss}^{\rm Lab} &= (\sqrt{s}, 0, 0, 0) - \sum_{i}^{n} p_{i}^{\rm Lab} \\
			&= p_{{\bf I}_a}^{\rm Lab} + p_{{\bf I}_b}^{\rm Lab},
	\end{split}\end{equation} 
	with $\sqrt{s}$ being the collision energy and $i$ being the index of each visible particle in the event. The recoil system ${\bf R}$ comprises all observable particles that are not assigned to the $\bf V$-systems. The initial state radiation (ISR) and imperfect object reconstruction of final states are the main sources of the $\bf R$ system in high energy lepton collisions~\cite{Yennie:1961ad}. 
 
\begin{figure}[th]
	\centering
	\includegraphics[width=0.9\linewidth]{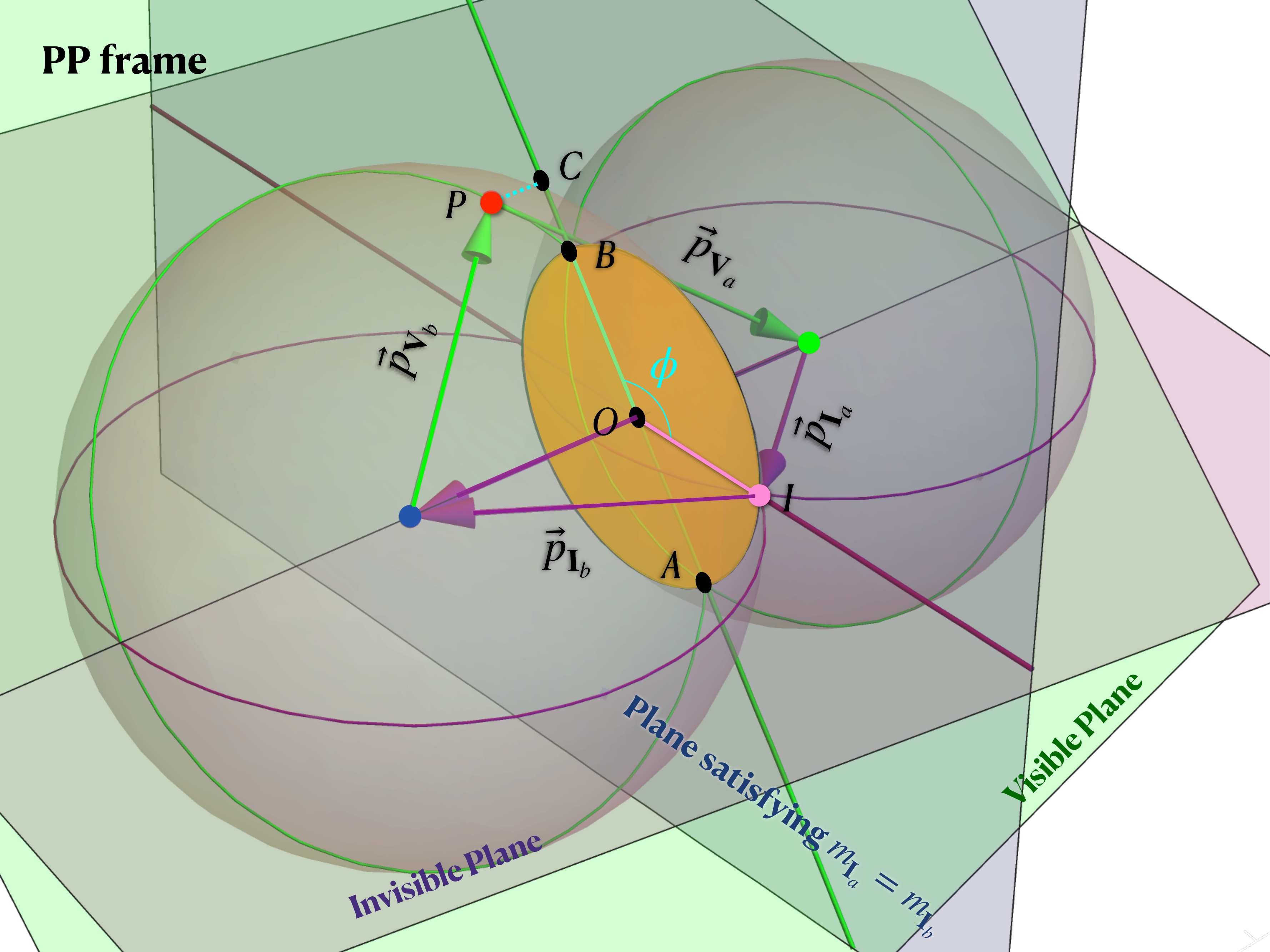}
	\caption{\label{fig2:momspace} The three-momentum vectors of semi-invisibly decaying particles shown in the {\bf PP} frame.}
\end{figure}

Boosting all the Lorentz vectors from the laboratory frame into the c.m. frame of the ${\bf PP}$ system, the three-momentums $\vec{p}_{{\bf V}_a}$, $\vec{p}_{{\bf I}_a}$, $\vec{p}_{{\bf V}_b}$ and $\vec{p}_{{\bf I}_b}$ can form two connected triangles with $\vec{p}_{\rm miss}$
\begin{equation}
	\vec{p}_{{\bf V}_a} + \vec{p}_{{\bf V}_b} + \vec{p}_{\rm miss} = 0, \quad 
	\vec{p}_{{\bf I}_a} + \vec{p}_{{\bf I}_b} - \vec{p}_{\rm miss} = 0,
\end{equation} 
in visible plane and invisible plane, as shown in FIG~\ref{fig2:momspace}. The angle between the two planes is denoted as $\phi$, and the vertex on the opposite side of $\vec{p}_{\rm miss}$ in the visible (invisible) triangle is marked as $P$ ($I$). Accordingly, the vector $\overrightarrow{P I~} = \vec{p}_{{\bf V}_a} + \vec{p}_{{\bf I}_a} = - \vec{p}_{{\bf V}_b} - \vec{p}_{{\bf I}_b}  $ is the three-momentum of the mother particle ${\bf P}$. 
\par The position of vertex $I$ is uncertain due to lack of knowledge of momentum distribution between two missing particles, which is nevertheless constrained by several conditions. Firstly, the real solution of $m_{\bf I}$ requires that the magnitude of the invisible three-momentum $\vec{p}_{{\bf I}_{a/b}}$ is limited by 
\begin{equation}\label{eq:LSP-posi}
\left| \vec{p}_{{\bf I}_{a/b}}\right| \leq E_{{\bf I}_{a/b}}=E_{{\bf P}_{a/b}}-E_{{\bf V}_{a/b}} = \frac{1}{2} E_{\bf PP} - E_{{\bf V}_{a/b}},
\end{equation}
where $E_{\bf PP} = E_{{\bf V}_a} + E_{{\bf V}_b}+E_{\rm miss} $ is the c.m. energy of the ${\bf PP}$ system. They imply the two spheres in FIG~\ref{fig2:momspace}, where the spherical shell corresponds the boundaries $m_{{\bf I}_{a/b}} = 0$. 
The particle $\bf I$ can also refer to a particle that undergoes invisible cascaded decay into a set of labeled particles ${\rm I}_{i}$, such as the neutralino $\tilde{\chi}_2^0 \to \tilde{\nu} \nu \to \nu \nu \tilde{\chi}_1^0$ case in supersymmetric theory. In this scenario, we have:
\begin{equation}
	m_{{\bf I}_{a/b}}^2 \equiv p_{{\bf I}_{a/b}}^2 = \left( \sum_{i} p_{{\rm I}_{i}}  \right)^2 \geq \sum_{i} m_{{\rm I}_{i}}^2 \geq 0. 
\end{equation} 
Hence, we can generally establish the validity of Eq.~(\ref{eq:LSP-posi}).
Secondly, in most scenarios, ${\bf I}_a$ and ${\bf I}_b$ are restricted to be the DM particles, which means their masses are identical\footnote{It should be emphasized that in more general cases, ${\bf I}_a$ and ${\bf I}_b$ may be the different particles with different masses. So, one should focus on the statistical distributions of our new variables. },
\begin{equation}
\begin{split}
	m_{{\bf I}}^2  = p_{{\bf I}_a}^{2} & = \left( \frac{1}{2} E_{\bf PP} - E_{{\bf V}_{a}} \right)^2 - \vec{p}_{{\bf I}_a}^{~2} \\
     = p_{{\bf I}_b}^{2} & = \left( \frac{1}{2} E_{\bf PP} - E_{{\bf V}_{b}} \right)^2 - \left(\vec{p}_{\rm miss} - \vec{p}_{{\bf I}_a} \right)^{2}. 
\end{split}
\end{equation}
Then one can get 
\begin{equation}
\begin{split}
	\vec{p}_{{\bf I}_a}  \cdot \vec{p}_{\rm miss}  = \frac{1}{2} & \left[ E_{{\bf V}_a}^2 - E_{{\bf V}_b}^2 \right. \\
	& \left. + E_{\bf PP}(E_{{\bf V}_b} - E_{{\bf V}_a} ) +  \left| \vec{p}_{\rm miss}\right|^2 \right], 
 \end{split}
\end{equation}
which says the vertex $I$ is located in a fixed plane vertical to $\vec{p}_{\rm miss}$, the blue plane in FIG~\ref{fig2:momspace}. 

\par Given the event topology in FIG~\ref{fig1}, the vertex $I$ locates in the interface between the spheres and the blue plane. Specifically, it lies on a flat, circular disc that is perpendicular to $\vec{p}_{\rm miss}$. 
There remains two unknown degrees of freedom, $\phi$ and $r$, where $r$ is the distance between the vertex $I$ and $\vec{p}_{\rm miss}$. For a given set $(\phi, r)$, we can obtain a set of masses $m_{\bf P}$ and $m_{\bf I}$, which determined respectively by the magnitude of vector $\overrightarrow{P I}$ and $\vec{p}_{{\bf I}_a}$. Hence, the geometric characteristics of the disc can be utilized to deduce the kinetic details of the event, which can be captured through three "reconstructed-mass" variables:
\begin{itemize}
	\item $m_{\rm LSP}^{\rm max}$: represents the maximum reconstructed value of $m_{\bf I}$, achieved by selecting the vertex $I$ at the center of the round disc $O$ in FIG~\ref{fig2:momspace}. Its value is given by
		\begin{equation}
		m_{\rm LSP}^{\rm max} = 
			\sqrt{ 
				E_{{\bf I}_a}^2 - \frac{1}{4 \left| \vec{p}_{\rm miss} \right|^2 }
				\left( \left| \vec{p}_{\rm miss} \right|^2 + E_{{\bf I}_a}^2 - E_{{\bf I}_b}^2 \right)^2 
			}.
		\end{equation}
	\item $m_{\rm RC}^{\rm min}$: represents the minimum reconstructed value of $m_{\bf P}$, which is obtained by setting $\left|\vec{p}_{{\bf P}a} \right|$ to its maximum value. This can be achieved by setting $\phi = \pi$ and maximizing the value of $r$ (placing vertex $I$ at point $A$ in FIG.~\ref{fig2:momspace}).
		\begin{equation}
			m_{\rm RC}^{\rm min} = \sqrt{\frac{E_{\bf PP}^2}{4} - \left| {PC} \right|^2 - \left(\left| {OC} \right| + \left| {OA} \right|  \right)^2  },
		\end{equation}
		where the point $C$ refers to the projection of vertex $P$ onto the blue plane, and $\left| {XY} \right|$ represents the distance between points $X$ and $Y$ in FIG.~\ref{fig2:momspace}. 
    The points $O, A, B,$ and $C$ lie along a straight line, which corresponds to the intersection of the visible plane and the blue plane. These distances can be determined geometrically,
    \begin{small}\begin{equation}\label{eq:trilength1}
	   \left| {PC} \right| = \frac{1}{2\left| \vec{p}_{\rm miss} \right| }
	   \Big(
	   	- E_{{\bf I}_a}^2 + E_{{\bf I}_b}^2 
		  + \left| \vec{p}_{{\bf V}_a} \right|^2 
		  - \left| \vec{p}_{{\bf V}_b} \right|^2 
		  \Big),
		  \end{equation}
		  \begin{equation}\begin{split}\label{eq:trilength2}
	   \left| {OC} \right| = \frac{1}{2\left| \vec{p}_{\rm miss} \right|} \bigg[
	    & 	-  \left| \vec{p}_{\rm miss} \right|^4 
	    	- \left(
	    		 \left|\vec{p}_{{\bf V}_a} \right|^2 
	    		- \left| \vec{p}_{{\bf V}_b} \right|^2 
	    	\right)^2 \\
 	    & 	+2 \left| \vec{p}_{\rm miss} \right|^2 
	    	\left(
	    		\left| \vec{p}_{{\bf V}_a} \right|^2 
	    		+ \left| \vec{p}_{{\bf V}_b} \right|^2
	    	\right) \bigg]^{\frac{1}{2}}, \\
	    \end{split}\end{equation}
		 \begin{equation}\begin{split}\label{eq:trilength3}
	    \left| {OA} \right| = \frac{1}{2\left| \vec{p}_{\rm miss} \right| } \bigg[
	    & 	- E_{{\bf I}_a}^4 
	    	- \left(
	    		E_{{\bf I}_b}^2
	    		-  \left| \vec{p}_{\rm miss} \right|^2 
	    	\right)^2 \\
	    &	+2 E_{{\bf I}_a}^2 
	    	\left(
			E_{{\bf I}_b}^2 
			+ \left| \vec{p}_{\rm miss} \right|^2 
		\right) \bigg]^{\frac{1}{2}}. 
    \end{split}\end{equation}\end{small}

    The $m_{\rm reconst}$ variable introduced in Ref.~\cite{Chen:2021omv} bears resemblance to $m_{\rm RC}^{\rm min}$. In this context, we present a more general analytical expression.
    
	\item $m_{\rm RC}^{\rm max}$ denotes the maximum reconstructed value of $m_{\bf P}$ obtained by selecting the minimum value of $\left|\vec{p}_{{\bf P}_a} \right|$. If point $C$ lies within the round disc, the minimal value of $\left|\vec{p}_{{\bf P}_a} \right|$ corresponds to the distance $\left| PC\right|$. Therefore, we have:
\begin{equation}
m_{\rm RC}^{\rm max} = \sqrt{\frac{E_{\bf PP}^2}{4} - \left| {PC} \right|^2 }.
\end{equation}

Alternatively, if point $C$ is outside the disc, implying the minimal value of $\left|\vec{p}_{{\bf P}_a} \right|$ is obtained by setting $\phi = 0$ and minimizing the value of $r$ (positioning vertex $I$ at point $B$), we have:
\begin{equation}
m_{\rm RC}^{\rm max} = \sqrt{\frac{E_{\bf PP}^2}{4} - \left| {PC} \right|^2 -\left(\left| {OC} \right| - \left| {OB} \right| \right)^2 },
\end{equation}
where $\left| {OB} \right| = \left| {OA} \right|$.    
\end{itemize}

From the definitions, the variables always follow the relations
\begin{equation}
\begin{split}
	0 \leq m_{\rm RC}^{\rm min} & \leq  m_{\bf P} \leq m_{\rm RC}^{\rm max} \leq \sqrt{s}\big/2 , \\
    0 \leq & m_{\bf I} \leq m_{\rm LSP}^{\rm max} \leq m_{\rm Recoil}\big/2. 
\end{split}
\end{equation}
In most scenarios, the invisible state $\bf I$ is typically assumed to be massless when determining $m_{\rm RC}^{\rm min}$ and $m_{\rm RC}^{\rm max}$. However, it is also possible to consider a non-zero prior $m_{\bf I}$ if necessary, thereby calculating $m_{\rm RC}^{\rm min}(m_{\bf I})$ and $m_{\rm RC}^{\rm max}(m_{\bf I})$ by replacing $E_{{\bf I}a}^2$ and $E_{{\bf I}_b}^2$ in Eq.~(\ref{eq:trilength3}) with $(E_{{\bf I}_a}^2 - m_{\bf I}^2)$ and $(E_{{\bf I}_b}^2 - m_{\bf I}^2)$ ], respectively.
For a given event, as $m_{\bf I}$ increases, the value of $m_{\rm RC}^{\rm min}\left(m_{\bf I}\right)$ increases, while $m_{\rm RC}^{\rm max}\left(m_{\bf I}\right)$ decreases. 
\vspace{.5cm}

\begin{figure}[ht]
	\makebox[\linewidth][c]{
	\includegraphics[width=0.53\linewidth]{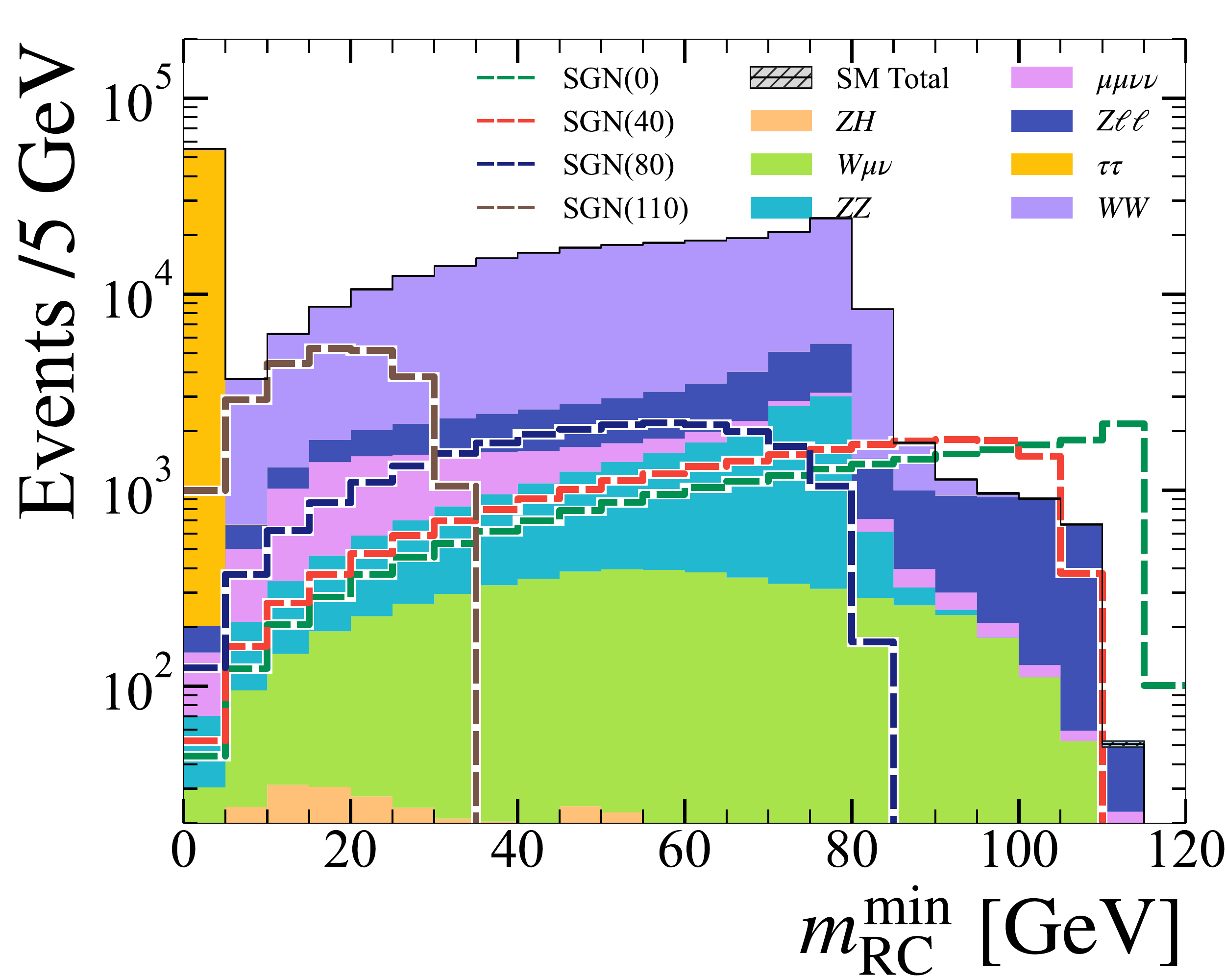}
	\includegraphics[width=0.53\linewidth]{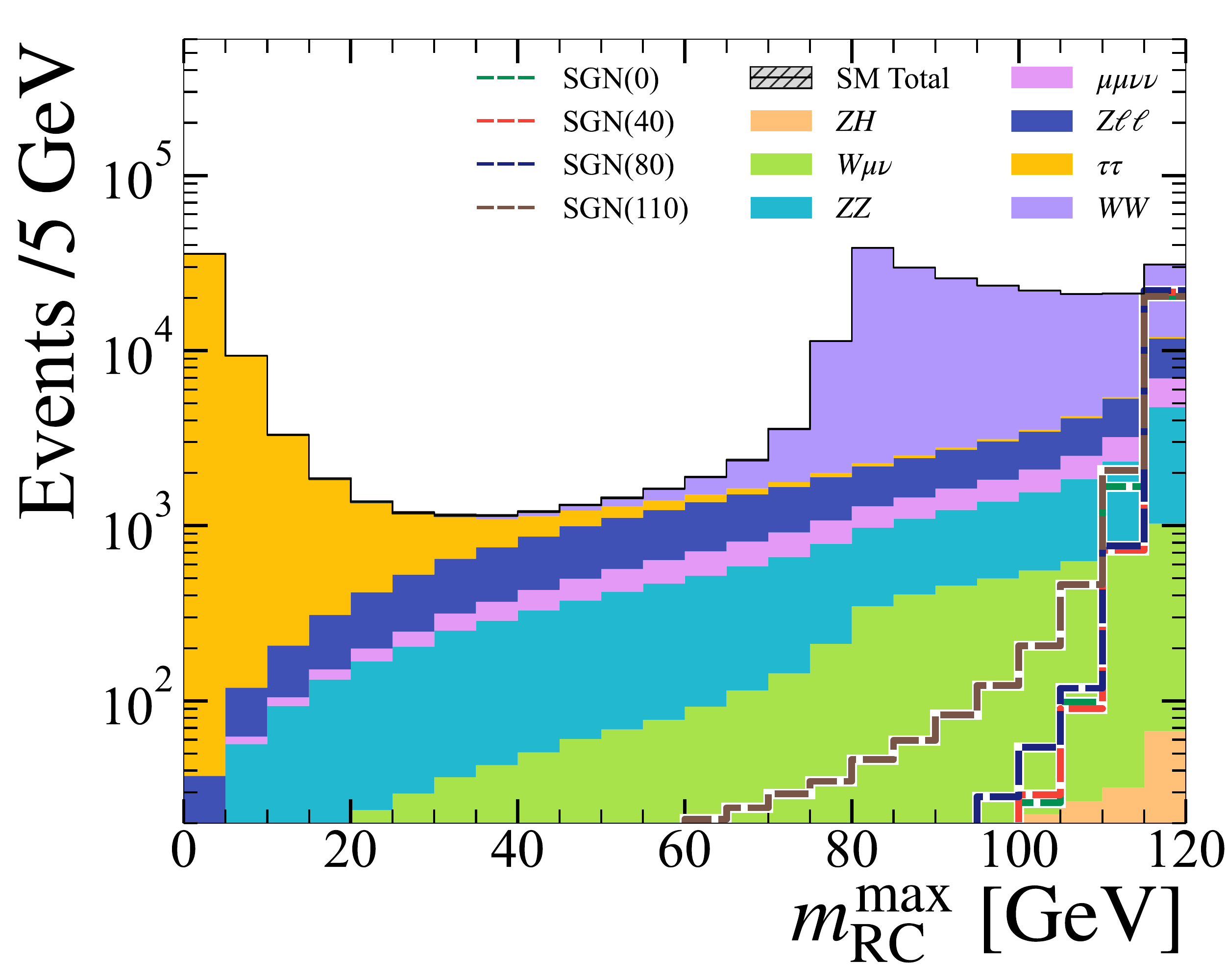}
    }
   	\makebox[\linewidth][c]{
	\includegraphics[width=0.53\linewidth]{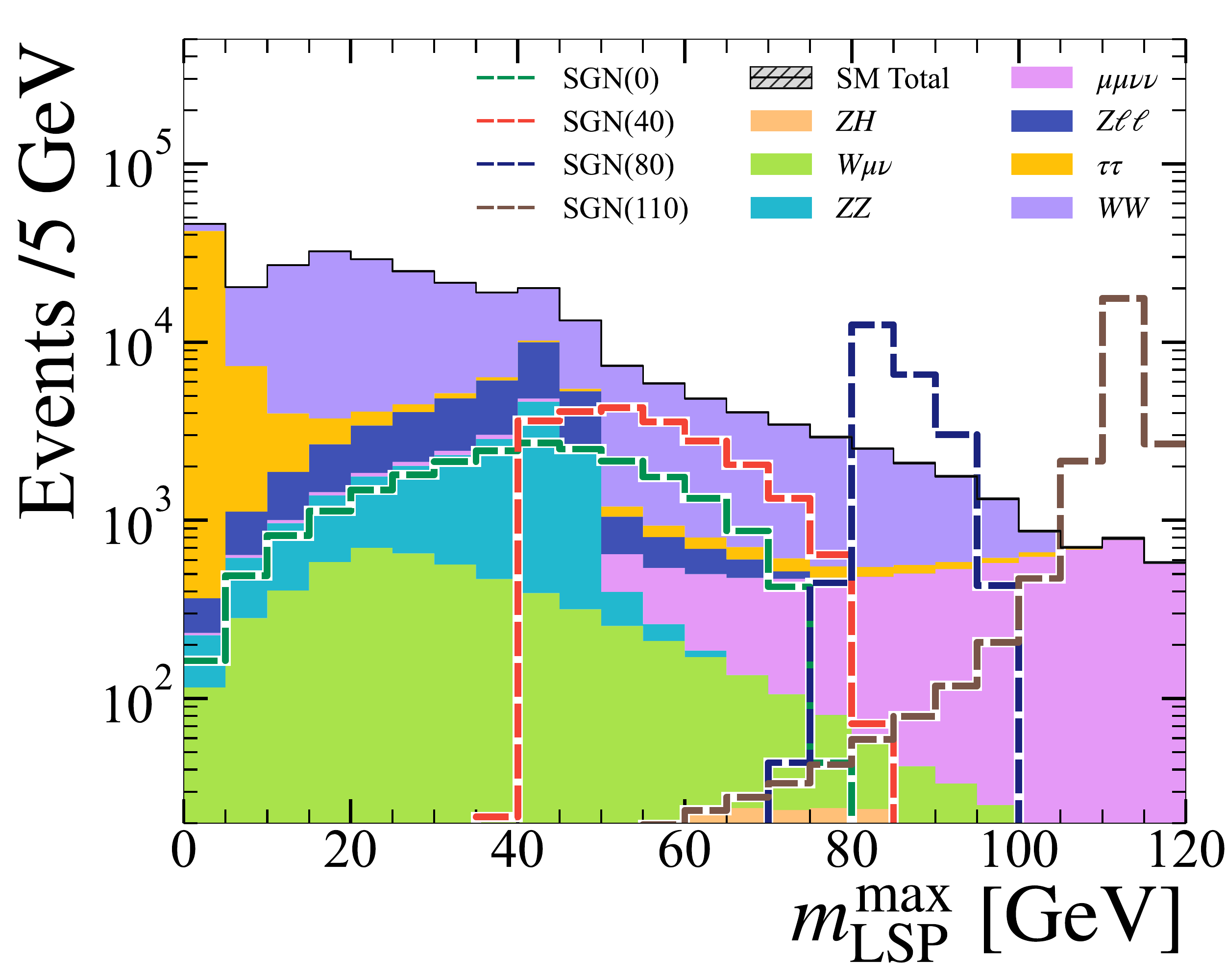}
	\includegraphics[width=0.53\linewidth]{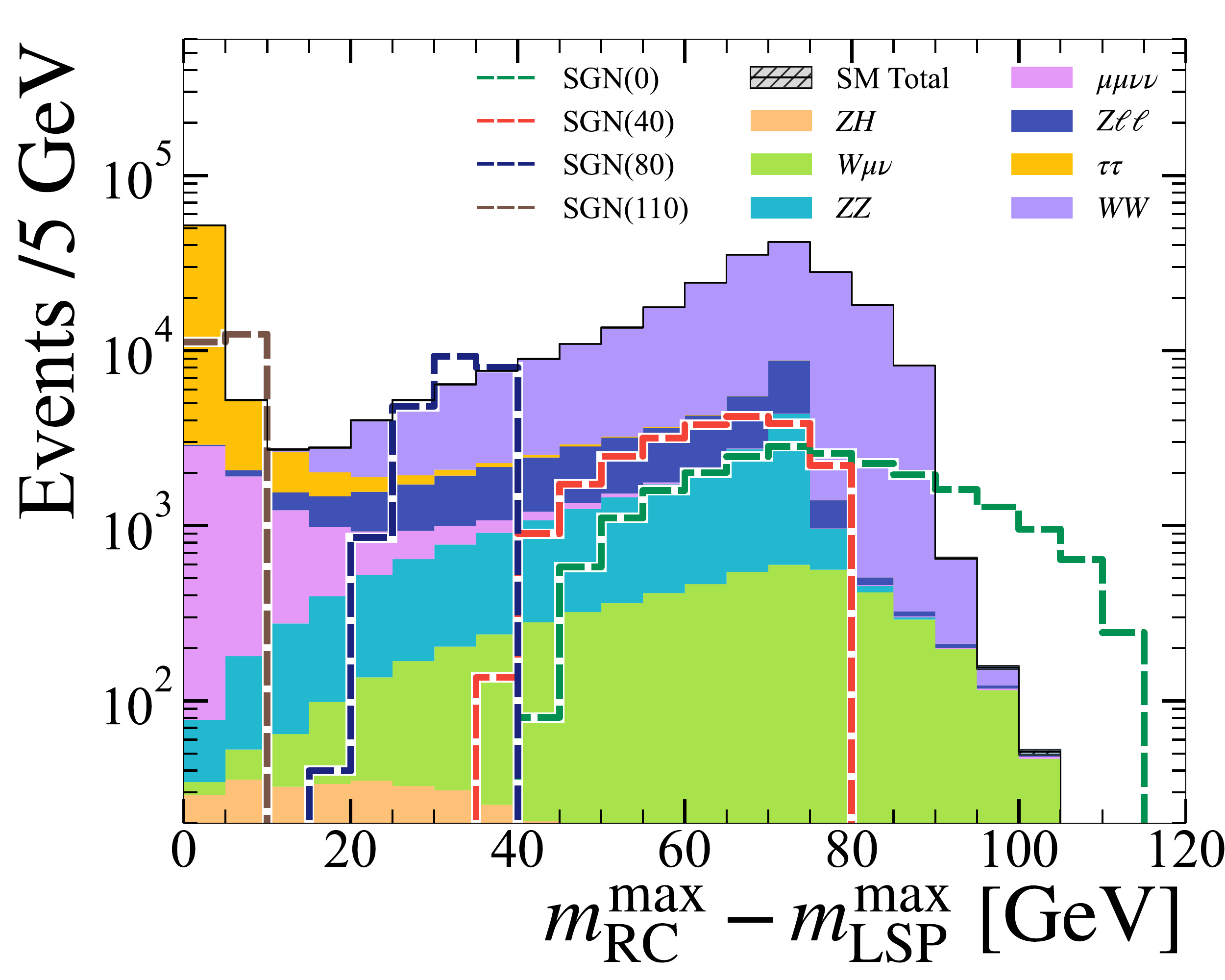}
    }
	\caption{\label{fig:cut} The distributions of the reconstructed masses at 240 GeV lepton collider of 5~ab$^{-1}$ after pre-selection criteria for the SM backgrounds and smuon pair production processes.}
\end{figure}
To be experimentally realistic, the main sources of uncertainties are:
\par 1. {\it Wrong assignment}. For a given event, the invisible weak initial state radiation which should be assigned into the $\bf R$ system is assigned into $\bf I$s, as shown in Eq.~(1), such as the invisible ISR $Z \to \nu\nu$. This can be neglected due to its tiny cross section. 
\par 2. {\it Initial/Final state radiation}. These two types of radiations not only smear the visible particle momenta but also provide a source for extra particles, such as soft jets and soft photons, in the events. 
\par 3. {\it The beam energy uncertainty}. The synchrotron radiation photon in the circular collider, the beamstrahlung process, etc., will cause the uncertainty of the collision energy $\sqrt{s}$. 
\par The excellent energy and momentum resolutions of the LCs allow for determination of all the inputs $p_{{\bf V}_a}^{\rm Lab}$, $p_{{\bf V}_b}^{\rm Lab}$, $p_{\rm miss}^{\rm Lab}$ and $p_{\bf R}^{\rm Lab}$ with good precision. However, all uncertainties listed above will contribute to the error of $\bf PP$ system total energy $E_{\bf PP}$, which further leads to the difference between the reconstruction and the reality. In practice, they all can be qualitatively estimated via the comparison between the Monte-Carlo simulation and the real experimental data, and can be eliminated by using some data-driven methods. At the current stage of our research, it is not realistic to analyze the complete sources of error and give numerical estimates. In the following, we take these effects into account as much as possible in the Monte Carlo simulations. 

{\bf Applications}~~~~ For illustration, we apply these new variables for the supersymmetry~\cite{Wang:2022rfd,Baer:2020kwz,Yang:2022qyz} search of smuon at $\sqrt{s} = 240~{\rm GeV}$ LC. We assume that the left-handed $\tilde{\mu}_L$ and right-handed $\tilde{\mu}_R$ are degenerated, the lightest neutralino $\tilde{\chi}_1^0$ is a nearly pure bino, and all other super particles are decoupled. Thus the branching ratio ${\rm Br}(\tilde{\mu} \to \mu \tilde{\chi}_1^0)=100\%$. The particles corresponding to $\bf P$, $\bf V$ and $\bf I$ in FIG.~\ref{fig1} are $\tilde{\mu}$, $\mu$ and $\tilde{\chi}_1^0$, respectively. This channel at CEPC has been investigated recently~\cite{Yuan:2022ykg, CEPCPhysicsStudyGroup:2022uwl}, using the azimuth angles $\Delta\phi(\mu^\pm, {\rm Recoil})$ and $\Delta\phi(\mu^+, \mu^-)$, the cone sizes $\Delta R(\mu^\pm, {\rm Recoil})$ and $\Delta R(\mu^+, \mu^-)$, the muon energies $E_\mu^{\pm}$, the sum of the transverse momentum of two muons ${\rm sum} P_{\rm T}$, the invariant mass $m_{\rm \mu\mu}$ and the $m_{\rm Recoil}$ variable. The results showed that the detection (discovery) limit of smuon mass can reach to 117 GeV (116 GeV), corresponding to a theoretical cross section of $11.1~{\rm fb}$ ($17.0~{\rm fb}$). The reconstructed mass variables we proposed can significantly improve these limits. 

\par In low energy supersymmetry (for some recent reviews, see, e.g., \cite{Wang:2022rfd,Baer:2020kwz,Yang:2022qyz}) , smuon pair production $e^+ e^- \to \tilde{\mu}^+ \tilde{\mu}^{-}$ at an $e^+ e^-$ collider occurs only via $\gamma$ or $Z$ exchange in the $s$-channel. The simplified model in the searching channel $ e^+ e^- \to \mu^+ \tilde{\chi}_1^0 \mu^- \tilde{\chi}_1^0 $ depends on the particle spectrum. Here $\tilde{\mu}$ contains both the left-handed $\tilde{\mu}_L$ and right-handed $\tilde{\mu}_R$, and their masses are degenerated which are labeled by $m_{\tilde{\mu}}$. The lightest neutralino $\tilde{\chi}_1^0$ is assumed to be a nearly pure bino, and the branching ratio ${\rm Br}(\tilde{\mu} \to \mu \tilde{\chi}_1^0)$ is set to 1.

\par In most cases, the cross section can be evaluated using the narrow width approximation 
\begin{equation}
	\sigma(\tilde{\mu}\tilde{\mu} \to  \mu^+ \tilde{\chi}_1^0 \mu^- \tilde{\chi}_1^0) = \sigma(\tilde{\mu} \tilde{\mu}) \times \left({\rm BR}(\tilde{\mu} \to \mu \tilde{\chi}_1^0)\right)^2 .
\end{equation}
At the CEPC, the electron and positron beams are unpolarized, and the cross section is written as~\cite{deCarlos:1995cx, Moortgat-Pick:2005jsx}
\begin{small}\begin{equation}\begin{split}
	\sigma(\tilde{\mu}\tilde{\mu}) &= \frac{(1-4m_{\tilde{\mu}}^2/s)^{3/2}}{24\pi s} \times \Bigg[e^4 \\ 
	& + \frac{e^2g^2}{2c_{W}^2} \left(\frac{1}{2}- 2 s_W^2 \right)^2 \frac{s(s - m_Z^2)}{(s-m_Z^2)^2 + \Gamma_Z^2 m_Z^2} \\
	& + \frac{g^4}{c_W^4}\left( \frac{1}{8}-\frac{1}{2} s_W^2 + s_W^4\right)^2 \frac{s^2}{(s-m_Z^2)^2 + \Gamma_Z^2 m_Z^2} \Bigg], 
\end{split}\end{equation}\end{small}
where $e$ is the electromagnetic coupling constant, $g$ is the weak force coupling constant, $s_W \equiv \sin{\theta_W}$ and $c_W \equiv \cos{\theta_W}$ with $\theta_W$ being the Weinberg angle, $m_Z$ is the $Z$-boson mass and $\Gamma_Z$ is the $Z$-boson width. For $m_{\tilde{\mu}} > \sqrt{s}/2$ or the parameter region where the narrow width approximation is invalid~\cite{Berdine:2007uv}, for example, the compressed region with $m_{\tilde{\chi}_1^0} \lesssim m_{\tilde{\mu}}$ and $m_{\tilde{\mu}} \lesssim \sqrt{s}/2$, the cross section is well described by a $2\to 3$ process, which suffers a phase space suppression
\begin{equation}\label{eq:xsect23}
	\sigma(\tilde{\mu}\tilde{\mu} \to  \mu^+ \tilde{\chi}_1^0 \mu^- \tilde{\chi}_1^0)= \sigma(\tilde{\mu} \mu \tilde{\chi}_1^0) \times {\rm BR}(\tilde{\mu} \to \mu \tilde{\chi}_1^0),
\end{equation}
where one smuon is produced off-shell and thus the smuon width should be correctly taken into account\footnote{A similar study of probing slepton like particle $S$ via the off-shell $2\to 3$ process can be found in Ref.~\cite{Liu:2021mhn}. }.

The dominated backgrounds are the direct $\tau^+\tau^-$ production and the processes involving electroweak bosons $W$/$Z$ and/or Higgs boson $h$ in the intermediate states. We separate the latter  into three classes: double-resonance background, single-resonance background and zero-resonance background, as defined in ~\cite{Cao:2018ywk}. The signal processes target the high smuon mass region at $m_{\tilde{\mu}}=115~{\rm GeV}$. Four LSP $\tilde{\chi}_1^0$ mass benchmark values are set at $0, 40, 80, 110~{\rm GeV}$, denoted as SGN(0), SGN(40), SGN(80), and SGN(110) respectively. 
Both signal and background processes listed in FIG.~\ref{fig:cut} are generated using \textsc{MadGraph5aMC@NLO}\cite{Alwall:2014hca,Frederix:2018nkq} and \textsc{PYTHIA8}~\cite{Bierlich:2022pfr}. \textsc{Rivet-3.1.6}~\cite{Buckley:2010ar} is used for detector simulation and event reconstruction. The decay width of smuon is evaluated by \textsc{SUSY-HIT}~\cite{Djouadi:2006bz}. 

Firstly, the events are required to contain exactly one opposite-sign~(OS) muon pair and no reconstructed jet objects, and the energy of both muons is required larger than 1 GeV and $|\eta| < 3.0$. A lower cut on the invariant mass of the recoil system, $m_{\rm Recoil} > 1~{\rm GeV}$, is used to reject direct $\mu\mu$ production. After this pre-selection, the distributions of $m_{\rm RC}^{\rm min}$, $m_{\rm RC}^{\rm max}$, $m_{\rm LSP}^{\rm max}$ and $\Delta m^{\rm max} \equiv m_{\rm RC}^{\rm max}- m_{\rm LSP}^{\rm max}$ are shown in FIG.~\ref{fig:cut}.

We see that for the processes of massless {\bf I}, such as $W^+W^-$ background or signal process of $m_{\tilde{\chi}_1^0}=0$, both $m_{\rm RC}^{\rm min}$ and $m_{\rm RC}^{\rm max}$ have an obvious peak truncating at $m_{\bf P}$. With $m_{\bf I}$ increasing, the distribution of $m_{\rm RC}^{\rm min}$ of signal process becomes flat and submerges into the backgrounds, while the peak of $m_{\rm LSP}^{\rm max}$ gets sharp and moves to high value. We also find that $\Delta m^{\rm max} \equiv m_{\rm RC}^{\rm max} - m_{\rm LSP}^{\rm max}$ can approximately represent the mass splitting between $\bf P$ and $\bf I$ for a heavy $\bf P$, which can be used for further reducing backgrounds.

We define three signal regions (SRs) of 12 bins aiming for different mass spectrum. Events with the invariant mass $m_{\mu\mu}$ and/or the recoil mass $m_{\rm Recoil}$ in $Z$-window of $10~{\rm GeV}$ is vetoed to suppress the processes with $Z$-boson resonance for all SRs. The cut $m_{\rm RC}^{\rm max} > 110~{\rm GeV}$ is used to focus on high smuon mass. The other detailed selections are listed in Table ~\ref{tab:sr}. The SRL, SRM and SRH categories are optimized respectively for the region of $m_{\tilde{\chi}_1^0}<40~{\rm GeV}$, $40~{\rm GeV}<m_{\tilde{\chi}_1^0}<80~{\rm GeV}$ and $m_{\tilde{\chi}_1^0}>80~{\rm GeV}$. The region SRL-02 is targeting one off-shell smuon events. 

The number of survived events in each SR for SM backgrounds and several signal processes at the 240 GeV LC of 5~ab$^{-1}$ are given in Table ~\ref{tab:sr}. The cross section for $m_{\tilde{\mu}}=115~{\rm GeV}$ is $23.9~{\rm fb}$. In the corresponding most sensitive SR, the acceptance rate is larger than 10\% for the signal, and the total cross section of background is suppressed to $0.05~{\rm fb} \sim 4~{\rm fb}$, implying a very large signal-to-background ratio.
The statistical sensitivity measurement is calculated in each SR for estimating the expected discovery significance~\cite{Cowan:2010js}. Here, the Poisson uncertainty of background and a conservative $5\%$ global systematic uncertainty are taken into consideration, and then the Asimov approximation of the detection and discovery significance $Z_A$ is given as  
\begin{equation}\begin{split}
	Z_A &= \left[ 2\left((s+b) \ln \left(\frac{(s+b)(b+\sigma_{b}^2)}{b^2 +(s+b)\sigma_b^2 } \right) \right. \right. \\
	&\quad - \left. \left. \frac{b^2}{\sigma_b^2}\ln\left( 1+ \frac{s\sigma_b^2}{b(b+\sigma_b^2)}\right) \right) \right]^{1/2}. 
\end{split}\end{equation}

\begin{table}[th]
\centering
\tabcolsep=2pt
\renewcommand{\arraystretch}{1.15} 
\resizebox{\linewidth}{!}{
\begin{tabular}{l| c cc cc cc c}
\hline\hline 
\multirow{1}{*}{SR} && SRL-01 && SRL-02 && SRM-01 && SRM-02  \\ \hline
 && $m_{\rm RC}^{\rm min}$ \textgreater 85 
 && \begin{tabular}[c]{@{}l@{}} 
 		$m_{\rm RC}^{\rm max}$ \textgreater 117 \\
 		$m_{\rm RC}^{\rm min}$ \textgreater 95 \\ 
 		$E_{\mu^\pm} \in [50, 70]$  
 	\end{tabular}
 && \begin{tabular}[c]{@{}l@{}} 
 		$m_{\rm RC}^{\rm min}>$  85\\ 
  		$m_{\rm LSP}^{\rm max} \in [40, 60]$ \\ 
 		$m_{\rm RC}^{\rm max}(40)>$  110 
 	\end{tabular}
 && \begin{tabular}[c]{@{}l@{}}
 		$m_{\rm LSP}^{\rm max} \in [50, 70]$ \\
		$m_{\rm RC}^{\rm max}(40)>$  110\\ 
 		$m_{\rm RC}^{\rm min}(40)>$  100 
 	\end{tabular}\\ \hline 
SM total  && 7532 $\pm$ 86  && 900 $\pm$ 30 && 2079 $\pm$ 45 && 970 $\pm$ 31\\ \hline 
SGN(0) && 38900 && 10900  && 6360 && 2270 \\
SGN(40) && 23800 && 1410 && 22100 && 10200  \\
SGN(80) && 0 && 0 && 0 && 0  \\
SGN(110) && 0 && 0 && 0 && 0 \\ 	
\hline
\hline
SR && SRM-03 && SRM-04 && SRM-05 && SRH-01  \\ \hline
 && \begin{tabular}[c]{@{}l@{}} 
 		$m_{\rm LSP}^{\rm max} \in [60, 80]$ \\
 		$m_{\rm RC}^{\rm min}(40)>$  95  
 	\end{tabular}
 && \begin{tabular}[c]{@{}l@{}} 
 		$m_{\rm LSP}^{\rm max} \in [70, 85]$ \\
 		$m_{\rm RC}^{\rm min}(70)>$  100 
 	\end{tabular}
 && \begin{tabular}[c]{@{}l@{}} 
 		$m_{\rm LSP}^{\rm max} \in [80, 95]$ \\ 
 		$m_{\rm RC}^{\rm max}(80)>$  105 
 	\end{tabular}
 && \begin{tabular}[c]{@{}l@{}}
 		$E_{\mu^\pm} \in [34, 44]$ \\
		$\Delta m^{\rm max} \in [35, 50]$
 	\end{tabular}\\ \hline 
SM total  && 912 $\pm$ 30  && 19941 $\pm$ 141 && 14039 $\pm$ 118 && 6412 $\pm$ 80\\ \hline 
SGN(0) && 59 && 468  && 0 && 49 \\
SGN(40) && 2460 && 4970 && 359 && 4340  \\
SGN(80) && 0 && 23900 && 74100 && 11600  \\
SGN(110) && 0 && 0 && 0 && 0 \\ 	
\hline
\hline
SR && SRH-02 && SRH-03 && SRH-04 && SRH-05  \\ \hline
 && \begin{tabular}[c]{@{}l@{}}
 		$E_{\mu^\pm} \in [28, 37]$ \\
		$\Delta m^{\rm max} \in [25, 40]$
 	\end{tabular}
 && \begin{tabular}[c]{@{}l@{}}
 		$E_{\mu^\pm} \in [22, 28]$ \\
		$\Delta m^{\rm max} \in [15, 30]$
 	\end{tabular}
 && \begin{tabular}[c]{@{}l@{}}
 		$E_{\mu^\pm} \in [15, 18]$ \\
		$\Delta m^{\rm max} \in [0, 20]$
 	\end{tabular}
 && \begin{tabular}[c]{@{}l@{}}
		$\Delta m^{\rm max} \in [0, 10]$
 	\end{tabular}\\ \hline 
SM total  && 5913 $\pm$ 76  && 3190 $\pm$ 56 && 786 $\pm$ 28 && 262 $\pm$ 12\\ \hline 
SGN(0) && 0 && 0  && 0 && 0 \\
SGN(40) && 5 && 0 && 0 && 0  \\
SGN(80) && 29200 && 11900 && 0 && 0  \\
SGN(110) && 0 && 0 && 0 && 112000 \\ 	
\hline\hline
\end{tabular}
}
\caption{\label{tab:sr} Summary of selection criteria of the smuon searches at the 240 GeV lepton collider, together with the background and signal expectations corresponding to an integrated luminosity of 5~ab$^{-1}$.}
\end{table}

\begin{figure}[th]
	\centering
	\includegraphics[width=0.95\linewidth]{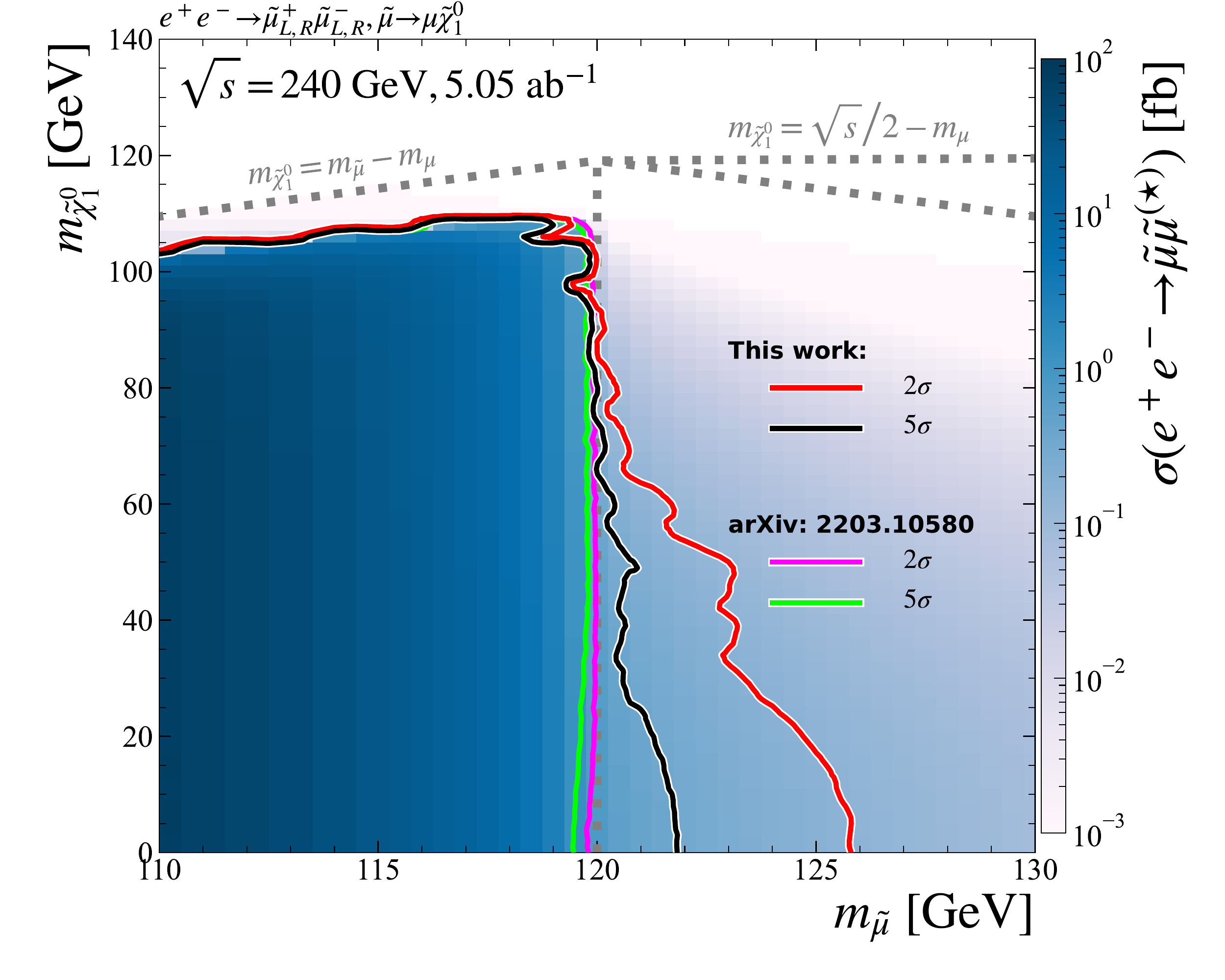}
	\caption{\label{fig:res} Projected exclusion and discovery reaches for direct smuon production. The left-handed and right-handed smuons are assumed to be degenerate, with the LSP being bino-like, while all other sparticles are decoupled. The shading colors are coded with the cross sections. }
\end{figure}
The expected detection and discovery sensitivities for the direct smuon production signals in the plane of $m_{\tilde{\mu}}$ versus $m_{\tilde{\chi}_1^0}$ is shown in FIG.~\ref{fig:res}. For each signal point, the SR with best $Z_A$ has been chosen in sensitivity mapping. We see that the detection (discovery) limit for a smuon can reach up to 126 GeV (122 GeV) for a massless $\tilde{\chi}_1^0$, corresponding to a smuon production cross section of $0.16~{\rm fb}$ ($0.34~{\rm fb}$). The limits break through the line of $\sqrt{s}\big/2$ and go into the off-shell region. We also recasted the analysis of Ref.~\cite{Yuan:2022ykg}, and their result shows that the detection limits of cross section can reach at about $3.2~{\rm fb}$ ($7.2~{\rm fb}$) for a massless $\tilde{\chi}_1^0$. Comparing with the recasted results of Ref.~\cite{Yuan:2022ykg}, we can find that the detection limits of the production cross section can be improved by one order of magnitude smaller. In Fig.~\ref{fig:res} we plot the corresponding contours for a more intuitive comparison. This kind of limits can also be obtained for other semi-invisible decaying particles, such as charginos which are pair produced at LCs. 

Here we adopt the cut-and-count method rather than more advanced technologies like machine learning~\cite{Guest:2018yhq, Abdughani:2019wuv, Feickert:2021ajf, Franceschini:2022vck}, in order to facilitate a comparison with previous results. The use of BDT~\cite{Roe:2004na, TMVA:2007ngy}, XGBoost~\cite{Chen_2016} and other techniques can further enhance the search capabilities, which we leave for future study. Additionally, these variables themselves can also be employed to the precision measurement of the SM particles, such as the tau lepton at BESIII and the $W$-boson at future LCs.

\begin{figure}[th]
	\makebox[\linewidth][c]{
	\includegraphics[width=0.5\linewidth]{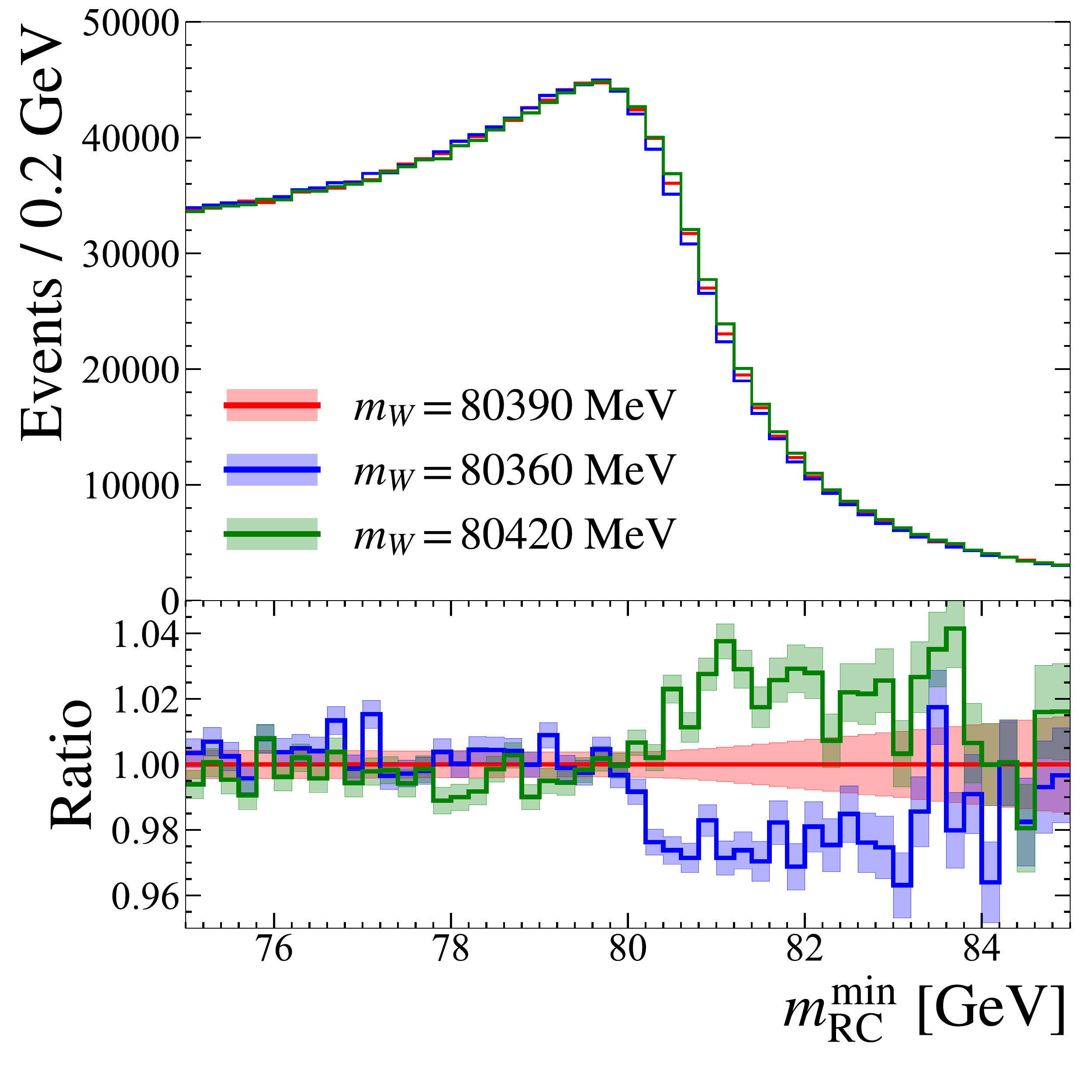}
	\includegraphics[width=0.5\linewidth]{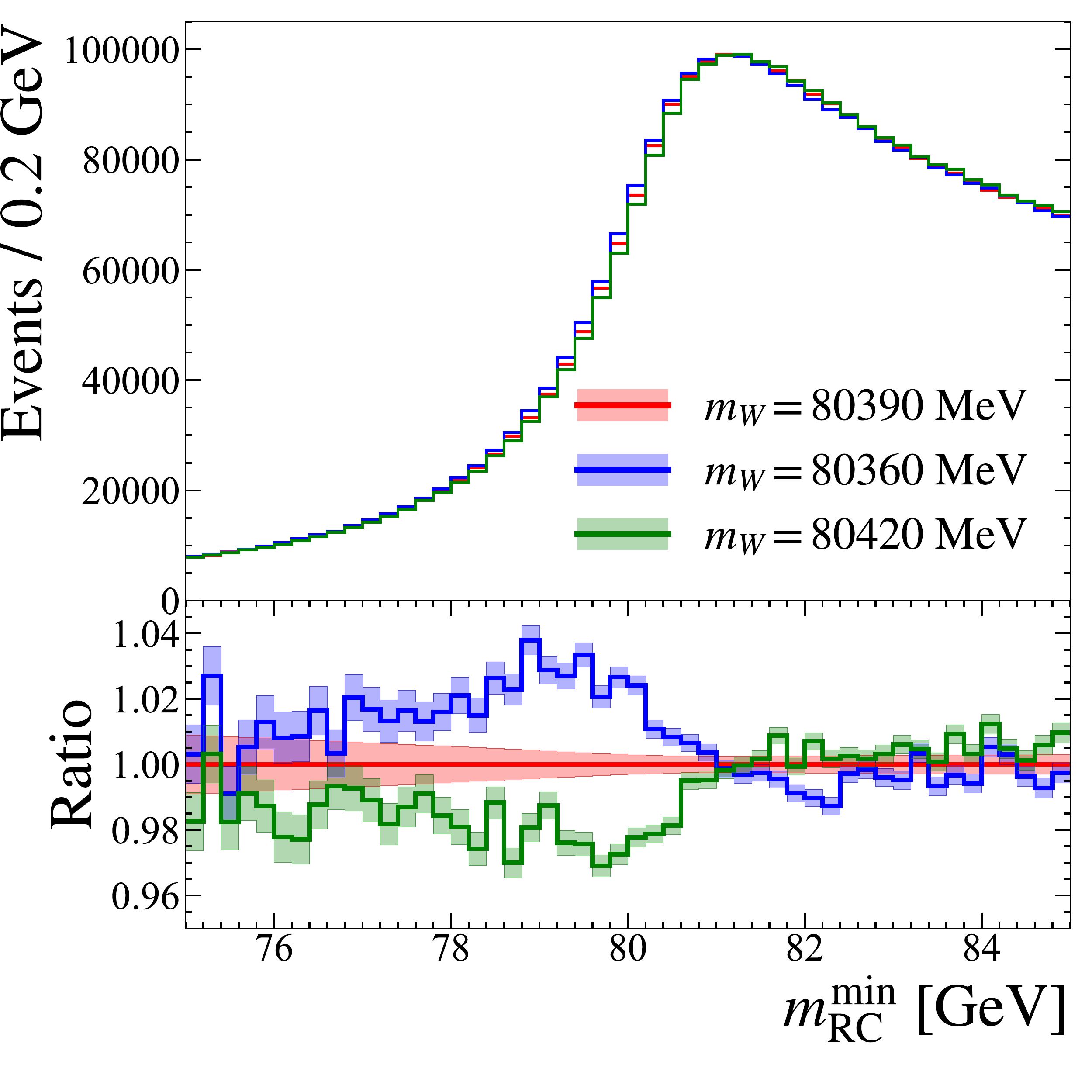}
	}
	\caption{\label{fig:wmassfit-distri} The distributions of $m_{\rm RC}^{\rm min}$ and $m_{\rm RC}^{\rm max}$ at the 240 GeV LC of 5~ab$^{-1}$ for $WW\to \ell \nu \ell^\prime \nu$ events. }
\end{figure}

Recently, a 7$\sigma$ deviation of $W$-boson mass from the SM prediction was reported ~\cite{CDF:2022hxs}. From Fig.~\ref{fig:wmassfit-distri}, one can find the distributions of $m_{\rm RC}^{\rm min}$ and $m_{\rm RC}^{\rm max}$ are sensitive to the $W$-boson mass, and their cutoff properties are similar to the $m_{\rm T}$ variable for $W^{\pm} \to \ell^{\pm} \nu$ events~\cite{Smith:1983aa,CDF:2004sns} and to the $m_{\rm T2}$ variable for $t\bar{t}$ events at hadron colliders, which can benefit the measurement of the $W$-boson mass at LCs.  
By assuming $m_W = 80390$ MeV and performing template fit at ILC/CEPC, the results of $W$-boson mass fitted via the distributions of $m_{\rm RC}^{\rm min}$ and $m_{\rm RC}^{\rm max}$ in $\ell\nu \ell^\prime \nu^\prime$ channel are 
\begin{equation}
    m_{W}^{\rm fit}=\left\{ 
    \begin{aligned}
    80391.98 &\pm 3.60~{\rm MeV}, &\text{by fitting }m_{\rm RC}^{\rm min};\\
    80389.42 &\pm 2.69~{\rm MeV}, &\text{by fitting }m_{\rm RC}^{\rm max}.
    \end{aligned} 
    \right. 
\end{equation}
The fitting accuracy is competitive to the measurement in the full-hadronic channel and in the semi-leptonic channel~\cite{CEPCPhysicsStudyGroup:2022uwl}, which is much better than the psedo-mass method (fitting the energy spectrum of the charged leptons)~\cite{Straessner:2004pw, OPAL:2002hhr} at LEP~\cite{DELPHI:2008avl, OPAL:2005rdt, L3:2005fft, ALEPH:2006cdc}.

\vspace{.5cm}
{\bf Summary}~~~~ 
We introduced a new set of variables to search for semi-invisible decaying particles pair-produced at LCs. By constructing the unknown degrees of freedom of the invisible states to a round disc in the center-of-mass frame, we defined $m_{\rm RC}^{\rm min}$, $m_{\rm RC}^{\rm max}$ and $m_{\rm LSP}^{\rm max}$ in a Lorentz-invariant way and gave the analytical expressions. They can be used to improve new particle searches and the precision measurement. With detailed Monte Carlo simulations for the 240 GeV LC of 5~ab$^{-1}$, we found that the significance of smuon searches can be improved by one order, and the precision of $W$-boson mass measurement via its full-leptonic decayed channel can be reduced to MeV level. The code of this work is available in \href{https://github.com/Buding820/mRC-variables}{GitHub page}~\cite{mRC:github}. 

{\em Acknowledgments.---}
The authors would like to thank Gang Li, Feng Lv, Kunlin Ran and Xuai Zhuang for helpful discussion.
This work was supported by the NNSFC (12105248, 11821505 and 12075300),  
by the Key Research Project of Henan Education Department for colleges and universities (21A140025),
by Peng-Huan-Wu Theoretical Physics Innovation Center (12047503),
by the CCEPP, and by the Key Research Program of the Chinese Academy of Sciences (XDPB15).

\bibliography{references}
\bibliographystyle{CitationStyle}

\end{document}